\documentclass[aps,prd,preprint,showpacs,superscriptaddress,groupedaddress,nofootinbib]{revtex4-1}
\usepackage{epsfig}
\usepackage{graphicx}
\usepackage{color}
\usepackage[normalem]{ulem}
\usepackage{amsmath}
\usepackage[breaklinks, plainpages=false, colorlinks=true, anchorcolor=cyan, linkcolor=red, citecolor=cyan, urlcolor=magenta, bookmarks=false]{hyperref}
\usepackage[caption=false]{subfig}
\begin{document}

\renewcommand{\thefigure}{\arabic{figure}}
\setcounter{figure}{0}

\def \univ{{\bf{n}}}
\def \unp{{\bf{p}}}
\def \uk{{\bf{k}}}

\bibliographystyle{apsrev}

\title{Gravitational Wave Searches with Pulsar Timing Arrays.\\ I:
  Cancellation of Clock and Ephemeris Noises} 
\author{Massimo Tinto}
\email{mtinto@ucsd.edu}
\affiliation{University of California San Diego,\\
  Center for Astrophysics and Space Sciences, \\
  9500 Gilman Dr, La Jolla, CA 92093, \\
  U.S.A.}  \date{\today}

\begin{abstract}
  We propose a data processing technique to cancel monopole and dipole
  noise sources (such as clock and ephemeris noises respectively) in
  pulsar timing array searches for gravitational radiation. These
  noises are the dominant sources of correlated timing fluctuations in
  the lower-part ($ \approx 10^{-9} - 10^{-8}$ Hz) of the
  gravitational wave band accessible by pulsar timing
  experiments. After deriving the expressions that reconstruct these
  noises from the timing data, we estimate the gravitational wave
  sensitivity of our proposed processing technique to single-source
  signals to be at least one order of magnitude higher than that
  achievable by directly processing the timing data from an equal-size
  array. Since arrays can generate pairs of clock and ephemeris-free
  timing combinations that are no longer affected by correlated
  noises, we implement with them the cross-correlation statistic to
  search for an isotropic stochastic gravitational wave background. We
  find the resulting optimal signal-to-noise ratio to be more than one
  order of magnitude larger than that obtainable by correlating pairs
  of timing data from arrays of equal size.
\end{abstract}

\maketitle

\section{Introduction}
\label{SecI}

The first LIGO \cite{LIGO} detection of a gravitational wave (GW)
signal from two medium mass Binary Black Hole (BBH) merger
\cite{GW150914}, followed by the observation of mergers from further
BBHs \cite{LV2,LV3,LV4} and one binary neutron star \cite{LV5}, mark
the beginning of GW astronomy. Because of seismic noise and limited
arm length issues, the lower part (below 10 Hz) of the GW spectrum
will only be accessible by space-based detectors such as the European
LISA mission \cite{LISA2017}, and pulsar timing arrays.

Pulsar timing GW experiments entail timing highly stable millisecond
pulsars in our own Galaxy. These experiments have been performed for
decades now, and they aim to detect gravitational radiation in the nHz
frequency band complementary to those accessible by ground and future
space-based GW detectors \cite{LISA2017,gLISA}. The basic principle
underlining the pulsar timing technique is the same as that of other
GW detector designs \cite{LIGO,VIRGO,DOPPLER,LISA2017}: to monitor the
frequency variations of a coherent electromagnetic signal exchanged by
two or more ``point particles'' separated in space. As a pulsar
continuously emits a series of radio pulses that are received at
Earth, a GW passing across the pulsar-Earth link introduces
fluctuations in the time-of-arrival (TOA) of the received
electromagnetic pulses. By comparing the pulses TOAs against those
predicted by a model, it is in principle possible to detect the
effects induced by any time-variable gravitational fields present,
such as the transverse-traceless metric curvature of a passing plane
GW train \cite{EW1975,Pulsar}. 

The frequency band in which the pulsar timing technique is most
sensitive to ranges from about $10^{-9}$ to $10^{-7}$ Hz, with the
lower-limit essentially determined by the overall duration of the
experiment ($10^{-9} \ {\rm Hz} \simeq 1/{\rm 30 \ years}$), and the
upper limit identified by the signal-to-noise ratio (SNR) of the
received radio pulses. To attempt observations of GWs in this way, it
is thus necessary to control, monitor and minimize the effects of
other sources of timing fluctuations, and, in the data analysis, to
use optimal algorithms based on the different characteristics of the
pulsar timing response to GWs (the signal) and to other sources of
timing fluctuations (the noise).

A quantitative analysis of the noise sources affecting millisecond
pulsar timing searches for gravitational radiation
\cite{JAT2011,Tiburzi2015,NANOGRAV1,NANOGRAV2} has shown that, in the
region ($10^{-9} - 10^{-7}$) Hz of the accessible frequency band, they
are due to:
\begin{enumerate}
\item finiteness of the SNR in the raw observations, resulting in what
  is usually referred to as {\it thermal noise} at the receiver;
\item uncertainties in solar system ephemeris, which are used to
  correct TOAs at the Earth to the barycenter of the solar
  system;
\item variation of the index of refraction in the interstellar and
  interplanetary plasma; 
\item intrinsic rotational stability of the pulsar, and
\item instability of a combination of the local clock and of the
  International Time Standard against which pulsars are timed, and
  noise in time transfer if the clock is not located at the
  observatory site.
\end{enumerate}
Although the timing fluctuations induced by some of these noises can
be in principle either reduced or calibrated out, the fundamental
noise-limiting sensitivity of pulsar timing experiments is imposed by
the timing-fluctuations inherent to the pulsar, the reference clocks
that control the TOA measurements, and the noise affecting the
ephemeris used for referring to the Solar System Barycenter (SSB) the
timing measurements performed at the observatory. The magnitudes of
these noises can be comparable to or larger \cite{JAT2011,Tiburzi2015}
than a GW stochastic background possibly present in the timing
data. For instance, clocks such as the Linear-Ion-Trapped-Standard
(LITS) (presently in-the-field state-of-the art atomic clock with
long-time-scale timing stability) \cite{LITS} would result in a
sinusoidal strain sensitivity of pulsar timing searches for GWs to a
level of about $10^{-15}$ after coherently integrating the data for a
period of $10$ years \cite{JAT2011}. Since the characteristic wave
amplitude associated with a super-massive black-hole-binaries
background is predicted to be of comparable magnitude at the frequency
$3 \times 10^{-9}$ Hz
\cite{Black1,Black2,Sesana,Schneider,NANOGRAV1,NANOGRAV2}, it is clear
that a single telescope will not be able to unambiguously detect such
a GW signal.

A method for statistically enhancing the SNR of pulsar timing
experiments to an isotropic background of GW was first proposed by
Hellings and Downs \cite{HD1983} and improved by Jenet {\it et al.}
\cite{JHLM2005}. This technique relies on cross-correlating pairs of
TOA residuals data taken with an array of highly-stable millisecond
pulsars.  Since the GW background is common to all timing residuals
while the noises affecting them may be uncorrelated, the
cross-correlation technique should enhance the strength of the GW
signal over that of the noises. In particular, it was shown that the
correlation of pulsar timing data as a function of their enclosed
angle has a characteristic signature that should enhance the
likelihood of detection by correlating timing data from a sufficiently
large ensemble of millisecond pulsars.

Although the correlation of the noise induced by the interstellar
medium and the intrinsic timing noise of each individual pulsar can be
disregarded since pulsars are generally widely separated on the sky,
some of the other above-mentioned noises are common to the array data
and result in non-zero correlations
\cite{Coles2011,Tiburzi2015,NANOGRAV1,NANOGRAV2}. This in turn may
prevent us from detecting the angular dependence of the Hellings-Downs
curve induced by a GW background. Among the various noise sources, the
instability of the International Time Standard against which pulsars
are timed and the noise associated with the SSB ephemeris are the most
important ones, showing strong correlations in the lower part of the
accessible frequency band \cite{NANOGRAV1,NANOGRAV2}. Although data
processing methods have been discussed in the literature to mitigate
the effects of correlated noises
\cite{Tiburzi2015,NANOGRAV1,NANOGRAV2} in pulsar timing arrays
searches for a GW stochastic background, here we propose a method to
cancel them. 

This article is organized as follows. In Sec. \ref{SecII} we present
the mathematical formulation of the problem after deriving the
transfer functions of the clock- and ephemeris noises in the TOA
residuals from an array of pulsars and noticing that they are
different from that of a GW signal. From these considerations we then
show that an array of $5$ pulsars allows us to: (i) identify a linear
combination of its TOA residuals that simultaneously cancels both
clock- and ephemeris noises; (ii) optimally reconstruct these
noises. In Sec. \ref{SecIII} we then estimate the GW sensitivity of
clock- and ephemeris-free combinations to single-source GW
signals. Since single-source searches are limited by clock and
ephemeris noises in the lower part of the accessible frequency band,
we estimate the sensitivity enhancement of our method over
single-pulsar searches to be at least one order of magnitude in this
part of the band. In Sec. \ref{SecIV} we then turn our attention to
searches for an isotropic stochastic GW background implemented by
cross-correlating pairs of clock- and ephemeris-free combinations
generated by arrays of $10$ or more pulsars. As the remaining noises
in some pairs of combinations are now uncorrelated
\cite{Tiburzi2015,NANOGRAV1,NANOGRAV2}, we derive the expressions of
the variance-covariance matrix associated with the correlation
statistic built with pairs of clock- and ephemeris-free data
combinations. We estimate that an array of $10$ pulsars, with its six
clock- and ephemeris-free combinations, can generate three pairs of
clock- and ephemeris-free combinations whose noises are uncorrelated
and achieve a sensitivity to an isotropic stochastic GW background
that is more than one order of magnitude better than that achievable
by cross-correlating TOA residuals from an equal size array.  In
Sec. \ref{SecV} we finally present our conclusions and considerations
about the technique we propose and the advantages it offers to the nHz
GW search efforts.

\section{Clock and Ephemeris Noise-Free Combinations}
\label{SecII}

Let us consider an array of $M$ pulsars timed by $M$ observatories
\footnote{The reason for using $M$ telescopes simultaneously tracking
  $M$ pulsars (where simultaneity is of course relative to the period
  of the waves searched for) rather than, for instance, $N$ telescopes
  tracking $M$ pulsars (with $N < M$) in a ``switching mode'' (i.e.,
  alternating between a subset of pulsars) is that the
  $M$-antennas-$M$-pulsars scenario will clearly show how our
  processing technique works.}. The SSB ephemeris, which are estimated
at the Jet Propulsion Laboratory (JPL) through analysis of tracking
data from interplanetary spacecraft \cite{KUCHYNKA2013243}, are used
to correct TOAs at the observatory to the SSB. The expression relating
the TOA at the SSB, $\hat{t}^{(i)}_{SSB}\ , \ i = 1, ... M$ to the TOA
at one of the $M$ observatories, $t^{(i)}_{obs} \ , \ i = 1, ... M$,
can be written in the following form \cite{LorimerKramer2005}
\begin{equation}
{\hat{t}^{(i)}}_{SSB} = t^{(i)}_{obs} - 
\frac{\hat{\bf{r}}^{(i)} (t) \cdot \hat{\univ}^{(i)}}{c} + \nu^{(i)} (t) \ , 
\label{Roemer}
\end{equation}
where the R\"omer delay has been shown explicitly, the symbol $\cdot$
represents the operation of scalar product between two vectors, $c$ is
the speed of light, and $\nu^{(i)} (t)$ includes all other
contributors to the difference between the Earth and SSB TOAs
\cite{LorimerKramer2005}. In Eq. (\ref{Roemer}),
${\hat{\bf{r}}}^{(i)} (t)$ and ${\hat{\univ}}^{(i)}$ denote the
position of the radio telescope $i$ and sky location of pulsar $i$
w.r.t. the SSB respectively, and the symbol $\hat{}$ on both
observables emphasizes that they are affected by errors.  By rewriting
them as sums of their ``true'' values,
(${\bf{r}}^{(i)} , {\univ}^{(i)}$), and their errors,
(${\bf{e}}^{(i)} (t), \Delta {\univ}^{(i)}$), Eq.(\ref{Roemer}) can be
rewritten in the following form (in which now the speed of light has
been taken to be equal to $1$)
\begin{equation}
{{\hat{t}}^{(i)}}_{SSB} \simeq t^{(i)}_{SSB} - 
{\bf{e}}^{(i)} (t) \cdot {\hat{\univ}}^{(i)}
- {\bf{r}}^{(i)} (t) \cdot \Delta {\univ}^{(i)}  + \nu^{(i)} (t) \ , 
\label{RoemerNew}
\end{equation}
where a term quadratic in the errors has been disregarded. Note that
the error vector ${\bf{e}}^{(i)} (t)$ can be decomposed into the sum
of two error terms, ${\bf{e}} (t) + {\bf{\eta}}^{(i)} (t)$, with the
first term representing the error of the center of the Earth relative
to the SSB and the second the error of the position of the observatory
$i$ relative to the center of the Earth. Since the position of
observatory $i$ with respect to the center of the Earth is known with
accuracy and precision that are orders of magnitude better than those
associated with the position of the center of the Earth relative to
the SSB \cite{Tiburzi2015}, in what follows our focus will be on the
SSB ephemeris used to convert TOAs from the Earth' s center to the
SSB. Under this assumption, Eq. (\ref{RoemerNew}) can be written in
the following form
\begin{equation}
{\hat{t}^{(i)}}_{SSB} \simeq t^{(i)}_{SSB} - {\bf{e}} (t) \cdot {\hat{\univ}}^{(i)}
- {\bf{r}}^{(i)} (t) \cdot \Delta {\univ}^{(i)} + \nu^{(i)} (t) \ .
\label{Roemer2}
\end{equation}
Note that the angular error associated with the sky location of the
pulsar w.r.t. SSB, $\Delta {\univ}^{(i)}$, can be as large as a few
tens of mas in both RA and DEC. This inaccuracy results in a
sinusoidal timing error of period one year and amplitude a few
microseconds \cite{Madisonetal2013}. Although the magnitude of such an
error is much larger than the other errors affecting the TOA
residuals, its well-defined frequency allows us to remove it
from the timing data and should not be regarded as a limiting
factor.

Let now $R^{(i)} (t) \ \ , i = 1, ...M$ be the TOA residual measured at
time $t$ by the $M$ radio telescopes timing $M$ pulsars
\footnote{Although TOA residuals from an array are generally sampled
  unevenly and at different times, by applying Fractional-Delay
  Filtering (FDF) \cite{FDF1,FDF2} to the timing data it is possible
  to reconstruct, with an exquisitely high accuracy, data points from
  the surrounding samples. As an example application of its use, FDF
  is integral part of the data processing technique used by LISA to
  digitally suppress (more than seven orders of magnitude) the laser
  noise by properly time-shifting and linearly combining the
  heterodyne measurements.}. From the noise-considerations made
earlier, the $M$-residuals can be described by the following
expressions
\begin{equation}
R^{(i)} (t)  =  H^{(i)} (t) \ + \  C (t) 
\ - \  {\bf{e}} (t) \cdot \hat{\univ}^{(i)}
\ + \ \zeta^{(i)} (t) \ \ , \ \ i =  1, ..., M \ ,
\label{Residuals}
\end{equation}
where the first term on the right-hand-side represents the
contribution from a possibly present GW signal
\cite{EW1975,Wahlquist87}, $C (t)$ is a monopole random process
associated with noises affecting all timing residuals at time $t$
(with the clock being probably the dominant one),
${\bf{e}} (t) \cdot \hat{\univ}^{(i)}$ is the ephemeris noise and
$\zeta^{(i)} (t)$ corresponds to the timing fluctuations due to all
other noise sources affecting the timing residual $i$. In what follows
we will assume the random processes $\zeta^{(i)} (t)$ to be of
zero-mean, and regard them as being uncorrelated to each other
\cite{Tiburzi2015}. 

Although the number of random processes we want to cancel are $4$ in
total, namely ($C (t), {\bf {e}} (t)$), the minimum number of TOA
residuals needed is actually $5$. This is because the exact removal of
the vector random process ${\bf {e}} (t)$ alone requires four timing
data. Since three directions to three pulsars are in general linearly
independent, it is possible to identify a linear combination of four
TOA residuals in which the resulting R\"omer terms add up to zero.

Before proceeding with the identification of the linear combinations
that simultaneously cancel the clock and ephemeris noises, we first
introduce an orthonormal basis (${\vec a}_1, {\vec a}_2, {\vec a}_3$)
centered on the SSB, and denote with
${\hat n}^{(i)}_j \ , \ i=1,\dots M \ , \ j = 1, 2, 3$ the components
of the unit vectors associated with the directions to the pulsars in
this coordinate system.

To identify the linear combinations that simultaneously cancel the
clock and ephemeris noises, we rewrite Eq.(\ref{Residuals}) in the
following matrix form
\begin{equation}
\begin{pmatrix}
    R^{(1)}  \\
    R^{(2)}  \\
    \vdots \\
    R^{(M)}  \\
\end{pmatrix}
=
\begin{pmatrix}
    1 & \hat{\univ}^{(1)} \\
    1 & \hat{\univ}^{(2)} \\
    \vdots \\
    1 & \hat{\univ}^{(M)}
\end{pmatrix}
\begin{pmatrix}
    C \\
    - \ {\bf {e}}
\end{pmatrix}
+
\begin{pmatrix}
    H^{(1)}  \\
    H^{(2)}  \\
    \vdots \\
    H^{(M)} 
\end{pmatrix}
+
\begin{pmatrix}
    \zeta^{(1)}  \\
    \zeta^{(2)}  \\
    \vdots \\
    \zeta^{(M)} \ .
\label{matrix}
\end{pmatrix}
\end{equation}
Since four or more unit vectors associated with the directions to the
pulsars are linearly dependent, we will assume three of them, say
($\hat{\univ}^{(1)}, \hat{\univ}^{(2)}, \hat{\univ}^{(3)}$), to be
linearly independent and use them as a new basis. This means that the
remaining $M - 3$ unit vectors can be written as linear combinations
of them as follows
\begin{equation}
\hat{\univ}^{(i)} = \sum_{j=1}^3 \alpha^{(i)}_j \hat{\univ}^{(j)} \ \
\ , \ \ \ i = 4,\dots M \ ,
\label{LinearDependence}
\end{equation}
where the $(M-3) \times 3$ matrix elements $\alpha^{(i)}_j$ are 
equal to
\begin{equation}
\alpha^{(i)}_j = \sum_{k=1}^3 {\hat n}^{(i)}_k ({\mathcal N}^{-1})^k_j \ \ , i =
4,\dots M \ \ , j = 1, 2, 3 \ ,
\end{equation}
and the $3 \times 3$ matrix ${\mathcal N}$ is given in terms of the
components of the three vectors ($\hat{\univ}^{(1)},
\hat{\univ}^{(2)}, \hat{\univ}^{(3)}$) by the following expression
\begin{equation}
{\mathcal N}
=
\begin{pmatrix}
    {\hat n}^{(1)}_1 & {\hat n}^{(1)}_2 & {\hat n}^{(1)}_3 \\
    {\hat n}^{(2)}_1 & {\hat n}^{(2)}_2 & {\hat n}^{(2)}_3 \\
    {\hat n}^{(3)}_1 & {\hat n}^{(3)}_2 & {\hat n}^{(3)}_3 \ .
\end{pmatrix}
\label{Transformation}
\end{equation}
After substituting Eq. (\ref{LinearDependence}) into the matrix
multiplying the vector $(C, - {\bf {e}})^T$ in Eq.(\ref{matrix}), the
problem becomes one of finding the generators of the Kernel
\cite{Lang} of this matrix. This means finding the vectors
${\vec \lambda} \equiv (\lambda_1, \lambda_2, \dots, \lambda_M)$ that, once
applied to the left of both sides of Eq.(\ref{matrix}), satisfy the
following homogeneous linear system
\begin{equation}
\begin{matrix}
(\lambda_1, \lambda_2, \dots, \lambda_M)
\end{matrix}
\begin{pmatrix}
    1 & \hat{\univ}^{(1)} \\
    1 & \hat{\univ}^{(2)} \\
    1 & \hat{\univ}^{(3)} \\
    \vdots \\
    1 & \sum_{j=1}^3 \alpha^{(i)}_j \hat{\univ}^{(j)} \\
    \vdots \\
    1 & \sum_{j=1}^3 \alpha^{(M)}_j \hat{\univ}^{(j)} \ .
\end{pmatrix}
= 0
\label{NullSpace1}
\end{equation}
The above equation translates in a corresponding homogeneous linear
system after noticing that a linear combination of three linearly
independent unit vectors is equal to zero {\it iff} the coefficients
of the combination are identically null. After some algebra we obtain
the following homogeneous linear system of $4$ equations in $M$
unknowns
\begin{equation}
\begin{pmatrix}
    1 & 1 & 1 & \dots & \dots & 1 \\
    1 & 0 & 0 & \alpha^{(4)}_1 & \alpha^{(5)}_1 & \dots \alpha^{(M)}_1 \\
    0 & 1 & 0 & \alpha^{(4)}_2 & \alpha^{(5)}_2 & \dots \alpha^{(M)}_2 \\
    0 & 0 & 1 & \alpha^{(4)}_3 & \alpha^{(5)}_3 & \dots \alpha^{(M)}_3 \\
\end{pmatrix}
\begin{pmatrix}
    \lambda_1 \\
    \lambda_2 \\
    \vdots \\
    \lambda_M
\label{NullSpace2}
\end{pmatrix}
= 0  \ .
\end{equation}
To find the generators of the null-space of the above $4 \times M$
matrix we have relied on the software {\it Mathematica}
\cite{Mathematica}. We have verified, for instance, that when $M = 4$
the Kernel is empty and the Image space \cite{Lang} has dimensionality
$4$. This means that with $4$ pulsars we can estimate the vector
$(C , - {\bf {e}})^T$. To this end, if we treat again the three unit
vectors ($\hat{\univ}^{(1)}, \hat{\univ}^{(2)}, \hat{\univ}^{(3)}$) as
basis, Eq.(\ref{matrix}) can be rewritten in the following form
\begin{equation}
\begin{pmatrix}
    R^{(1)}  \\
    R^{(2)}  \\
    R^{(3)}  \\
    R^{(4)}  
\end{pmatrix}
=
\begin{pmatrix}
    1 & 1 & 0 & 0 \\
    1 & 0 & 1 & 0 \\
    1 & 0 & 0 & 1 \\
    1 & \alpha^{(4)}_1 & \alpha^{(4)}_2 & \alpha^{(4)}_3
\end{pmatrix}
\begin{pmatrix}
    C \\
    - \ {\bf {e}} \cdot \hat{\univ}^{(1)} \\
    - \ {\bf {e}} \cdot \hat{\univ}^{(2)} \\
    - \ {\bf {e}} \cdot \hat{\univ}^{(3)} 
\end{pmatrix}
+
\begin{pmatrix}
    H^{(1)}  \\
    H^{(2)}  \\
    H^{(3)}  \\
    H^{(4)}  
\end{pmatrix}
+
\begin{pmatrix}
    \zeta^{(1)}  \\
    \zeta^{(2)}  \\
    \zeta^{(3)}  \\
    \zeta^{(4)}  \\
\label{matrix2}
\end{pmatrix}
\ .
\end{equation}
The inverse of the matrix multiplying the vector
$(C, - \ {\bf {e}} \cdot \hat{\univ}^{(1)} , - \ {\bf {e}} \cdot
\hat{\univ}^{(2)} , - \ {\bf {e}} \cdot \hat{\univ}^{(3)})^T$
in Eq. (\ref{matrix2}) can easily be derived and its expression is
equal to
\begin{equation}
A \equiv \frac{1}{\alpha^{(4)}_1 + \alpha^{(4)}_2 + \alpha^{(4)}_3 -
  1} \ 
\begin{pmatrix}
    \alpha^{(4)}_1 & \alpha^{(4)}_2 & \alpha^{(4)}_3 & - 1 \\
    \alpha^{(4)}_2 + \alpha^{(4)}_3 -1 & - \alpha^{(4)}_2 & - \alpha^{(4)}_3 & 1 \\
    - \alpha^{(4)}_1 & \alpha^{(4)}_1 + \alpha^{(4)}_3 - 1 &  - \alpha^{(4)}_3 &  1 \\
    - \alpha^{(4)}_1 & - \alpha^{(4)}_2 & \alpha^{(4)}_1 +
    \alpha^{(4)}_2 - 1 &  1  
\end{pmatrix}
\ .
\label{CEMatrix}
\end{equation}
By left-applying the matrix $A$ to both sides of Eq.(\ref{matrix2}),
we obtain the following estimate for the clock and ephemeris noises
\begin{equation}
A  
\begin{pmatrix}
    R^{(1)}  \\
    R^{(2)}  \\
    R^{(3)}  \\
    R^{(4)} 
\end{pmatrix}
=
\begin{pmatrix}
    C \\
   - \ {\bf {e}} \cdot \hat{\univ}^{(1)} \\
   - \ {\bf {e}} \cdot \hat{\univ}^{(2)} \\
   - \ {\bf {e}} \cdot \hat{\univ}^{(3)}
\end{pmatrix}
+
A 
\begin{pmatrix}
    H^{(1)}  \\
    H^{(2)}  \\
    H^{(3)}  \\
    H^{(4)} 
\end{pmatrix}
+ A 
\begin{pmatrix}
    \zeta^{(1)}  \\
    \zeta^{(2)}  \\
    \zeta^{(3)}  \\
    \zeta^{(4)} 
\label{matrix3}
\end{pmatrix}
\ .
\end{equation}
Finally, the expressions for the components of the ephemeris noise
w.r.t. the orthonormal basis (${\vec a}_1, {\vec a}_2, {\vec a}_3$)
can be obtained by applying the matrix ${\mathcal N}^{-1}$ to the left
of the reconstructed components of the ephemeris given by
Eq.(\ref{matrix3}).

The configurations with $M > 4$ have a Kernel of dimension equal to
$M-4$ and an Image that is $4$-dimensional. This means that with
$M > 4$ pulsars it is possible to reconstruct the clock and ephemeris
noises in four different and independent ways. Although the noises
affecting these reconstructions may be correlated, there should exist
an {\underline {optimal}} combination of the reconstructed clock and
ephemeris noises that improves upon each individual reconstruction. We
will analyze the problem of optimally reconstructing the clock and
ephemeris noises from linearly combining pulsar timing data in a
future publication as the focus of this article is on searches for GWs
with clock- and ephemeris-free combinations of timing data.

\begin{figure}[htp]
\includegraphics[clip=true,angle=0,width=1.0\textwidth]{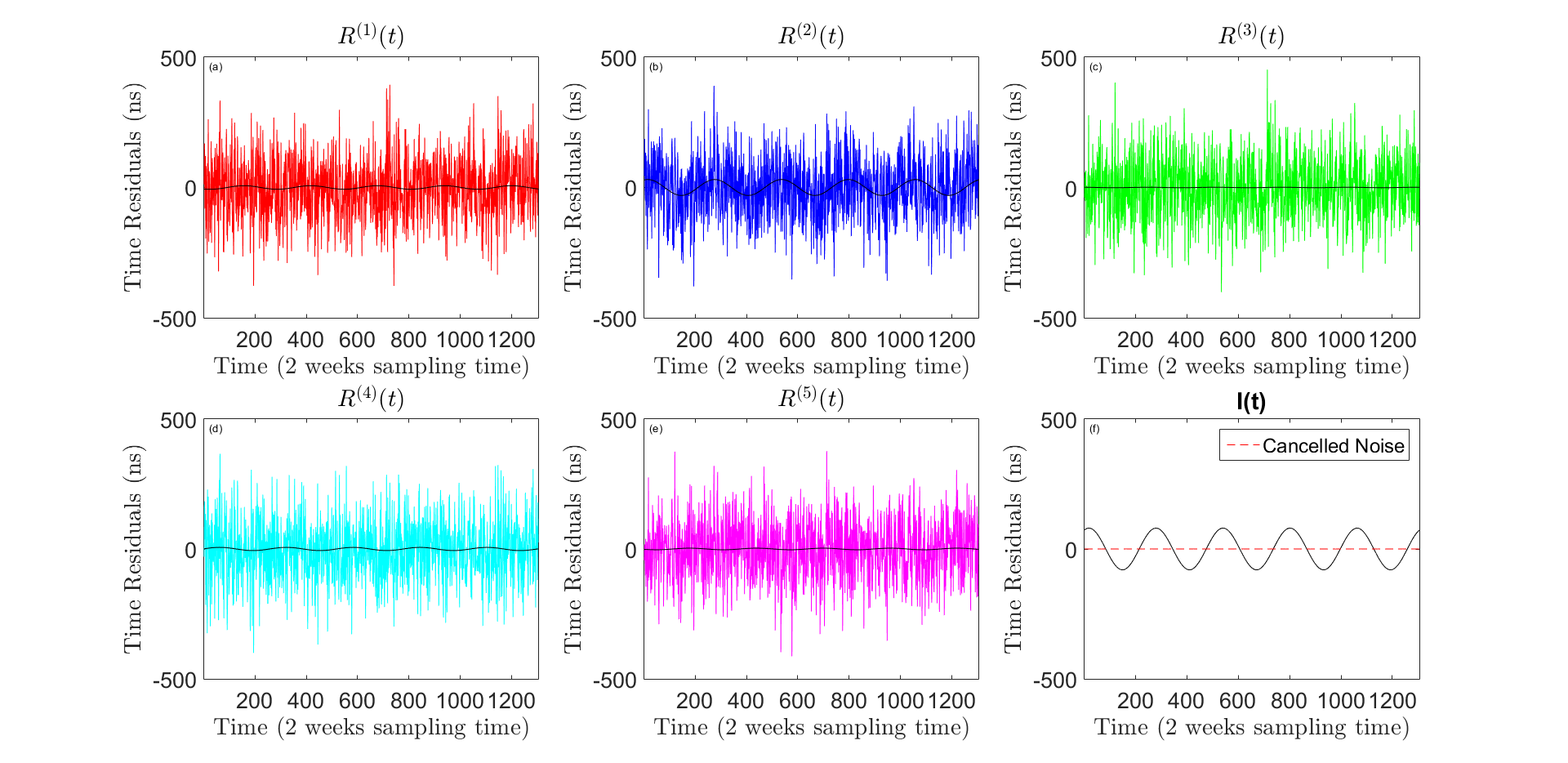}
\caption{\label{fig1} Simulation of the combination $I(t)$. The
  inserts (a) through (e) show five timing residuals containing a GW
  signal (black line) and noises from clock and ephemeris. The GW
  signal was assumed to be sinusoidal with frequency equal to
  $3.3 \times 10^{-9}$ Hz, and with its two strain polarization
  amplitudes equal to $6.0 \times 10^{-16}$ (each corresponding to
  about $30$ ns at the chosen frequency). The clock and ephemeris noises
  were taken to be Gaussian distributed random processes of zero-mean
  and r.m.s. amplitudes equal to $100$ ns (clock) and $100/\sqrt{3}$
  ns for each of the three components of the ephemeris noise. The
  pulsars were assumed to be at a distance of 1 kpc and at some random
  directions in the sky; insert (f) shows $I(t)$ after applying the
  noise-canceling algorithm.}
\end{figure}

In the specific case $M = 5$, the components of the single generator
of the Kernel, ${\vec \lambda}$, are equal to the following expressions
(up to a scaling constant chosen to be equal to $1$)
\begin{eqnarray}
\lambda_1 & = & (\alpha^{(5)}_2 + \alpha^{(5)}_3 - 1) \alpha^{(4)}_1 + (1 - \alpha^{(4)}_2 -
                \alpha^{(4)}_3) \alpha^{(5)}_1 \ ,
\nonumber
\\
\lambda_2 & = & (\alpha^{(5)}_3 + \alpha^{(5)}_1 - 1) \alpha^{(4)}_2 + (1 - \alpha^{(4)}_3 -
                \alpha^{(4)}_1) \alpha^{(5)}_2 \ ,
\nonumber
\\
\lambda_3 & = & (\alpha^{(5)}_1 + \alpha^{(5)}_2 -1) \alpha^{(4)}_3 + (1 - \alpha^{(4)}_1 -
                \alpha^{(4)}_2) \alpha^{(5)}_3 \ ,
\nonumber
\\
\lambda_4 & = & (1 - \alpha^{(5)}_1 - \alpha^{(5)}_2 - \alpha^{(5)}_3) \ ,
\nonumber
\\
\lambda_5 & = & (\alpha^{(4)}_1 + \alpha^{(4)}_2 + \alpha^{(4)}_3 - 1) \ .
\label{lambdas}
\end{eqnarray}
To exemplify the efficacy of the time-series
$I(t) \equiv \sum_{i=1}^5 \lambda_i R^{(i)} (t)$ in canceling clock
and ephemeris noises with five pulsars, we have numerically
synthesized five timing residuals $R^{(r)} (t) \ , \ r=1,...5$ by
generating clock and ephemeris noises through the use of a Gaussian
random number generator. Both noises have been assumed to have
zero-mean and a root-mean-squared (r.m.s.) error equal to $100$ ns and
$100/\sqrt{3}$ ns for each of the three components of the ephemeris
noise respectively. In addition, we have added a GW signal
characterized by two sinusoidal polarization components with strain
amplitudes each equal to $6.0 \times 10^{-16}$ and frequency equal to
$3.3 \times 10^{-9}$ Hz. The wave's propagation direction and the
directions to the five pulsars were randomly selected, while the
contribution from all other noises (denoted
$\zeta^{(r)} \ \ \ , r=1,...5$ in the above equations) affecting the
measured timing residuals were not included to visually emphasize the
clock and ephemeris noise cancellation in $I(t)$. Figure \ref{fig1}
shows the five residuals (insert (a) through (e)) simulated over a
period of $50$ years and sampled every two weeks. The black-colored
lines represent the response of each timing residual to the above GW
signal; insert (f) shows the effectiveness of the noise-canceling
algorithm by synthesizing the TOA residual combination $I(t)$.

Although our analysis treats the pulsars' sky locations as constants,
it is clear that the linear combination $I(t)$ works with pulsars that
may move across the sky. $I(t)$ should in fact be regarded as an
example application of the more general data processing technique
called {\it Time-Delay Interferometry} (TDI)
\cite{TintoDhurandhar2014}. By properly time-shifting and linearly
combining data measured by a network of GW detectors, TDI provides the
mathematical framework for deriving new time-series that are
unaffected by correlated noises while retaining sensitivity to GWs.

\section{Sensitivity to Single-Source Signals} 
\label{SecIII}

To quantify the advantages brought by our data processing technique,
we will first derive the expression of the sensitivity to individual
GW signals when $M=5$, and compare it against that of a single pulsar.
To this end, let us first rewrite the expression of $I(t)$ in the
following form
\begin{equation}
  I(t) \equiv H(t) + N(t) = \sum_{r=1}^5 \lambda_r H^{(r)} + \sum_{r=1}^5 \lambda_r \zeta^{(r)} (t) \ ,
\label{ClokEphemerisfree}
\end{equation}
where $H(t)$ corresponds to the first term on the right-hand-side (the
GW signal in $I$), and $N(t)$ to the second (the noise in $I$). To
estimate the sensitivity of $I$ to individual GW signals, we first
derive the expression of the GW signal power averaged over sources
randomly distributed on the sky and polarization states
\begin{equation}
  \langle |{\widetilde{H}} (f)|^2 \rangle =
  \sum_{i,j=1}^5 \lambda_i \lambda_j 
  \langle {\widetilde{H}}^{(i)} (f) {\widetilde{H}}^{*(j)} (f) \rangle = \sum_{i,j=1}^5 \lambda_i \lambda_j \Theta^{ij} h^2(f) \ ,
\label{H2}
\end{equation} 
where the angle-brackets denote the averaging operation over the
celestial sphere and wave's polarization states, $\widetilde{}$
represents the operation of Fourier transform and $^*$ that of complex
conjugation. In Eq. (\ref{H2}) $\Theta^{ij}$ is the Hellings-Downs
\cite{HD1983} correlation function of the angle enclosed by the
directions to the two pulsars ($i,j$), and
$h^2 (f) \equiv \langle |{\widetilde H}^{(i)} (f)|^2 \rangle \ , \
i=1,...5$,
is the averaged GW power in each timing residual. This can be written
as $h^2(f) = \rho^2(f) \ h^2_0(f)$, where $h_0(f)$ is the Fourier
transform of the wave amplitude and $\rho(f)$ is the resulting wave's
r.m.s. transfer function to the pulsar response
\cite{EW1975,AET1999,JAT2011}. 

Since the noises $\zeta^{(r)} \ , r=1,...5$ are uncorrelated and each
can be characterized by its own one-sided power spectral density,
$P_{\zeta^{(r)}}(f)$, the expression of the one-sided power spectral
density of the noise $N$, $P_N (f)$, is equal to
\begin{equation}
P_N (f) = \sum_{r=1}^5 \lambda^2_r P_{\zeta^{(r)}}(f) \ .
\label{Spectrum}
\end{equation}
The GW sensitivity of the $I(t)$ combination, $\gamma (f)$, defined as
the ratio between the square-root of its noise spectrum, $P_N(f)$, and
the r.m.s. transfer function of its GW response,
$\sqrt{\sum_{i,j=1}^5 \lambda_i \lambda_j \Theta^{ij} \rho^2 }$, is
then equal to \cite{Thorne1987,AET1999}
\begin{equation}
  \gamma (f) = \sqrt{\frac{\sum_{r=1}^5 \lambda^2_r P_{\zeta^{(r)}}(f)}
    {\sum_{i,j=1}^5 \lambda_i \lambda_j \Theta^{ij} \rho^2(f)}} \ .
\label{Sensitivity1}
\end{equation}

\begin{figure}[htp]
\includegraphics[clip=true,angle=0,width=1.0\textwidth]{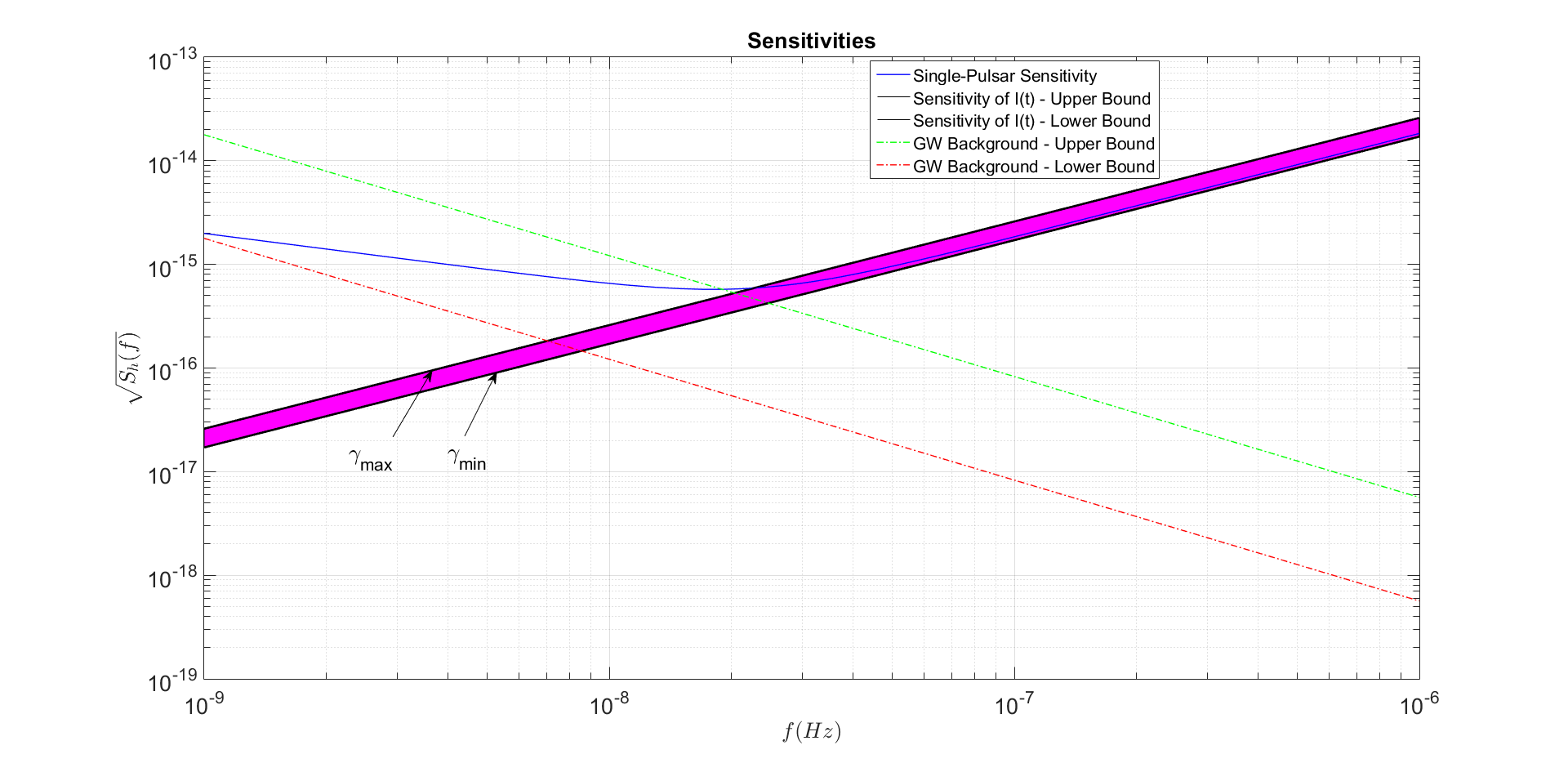}
\caption{\label{fig2} Gravitational wave sensitivity expressed as the
  ratio between the square-root of the one-sided power spectral
  density of the noise and the r.m.s. of the GW signal transfer
  function. The pulsars have been assumed to be at equal distances
  from Earth (1 kpc). The sensitivity of a single-telescope experiment
  (blue curve), as well as the current estimates of the upper- and
  lower-bounds (green and red line respectively) on the amplitude
  radiated by an ensemble of super-massive black-hole binaries are
  included \cite{Sesana,NANOGRAV1}. Note the narrowness of the
  sensitivity region (bounded by the functions
  $\gamma_{max}, \gamma_{min}$) within which the sensitivity of the
  data combination $I(t)$ is defined. The GW sensitivity gains of
  $I(t)$ over that of a single-pulsar experiment are evident in the
  frequency interval ($10^{-9} - 2.0 \times 10^{-8}$) Hz. The
  sensitivity curves presume adequate calibration of timing
  fluctuations from intergalactic and interplanetary plasma and
  negligible pulsars spin noises. See text for details.}
\end{figure}

To get some insights about $\gamma (f)$, let us consider the case of
timing residual noises $\zeta^{(r)}$ being characterized by the same
spectrum, i.e.
$P_\zeta (f) \equiv P_{\zeta^{(r)}}(f) \ \ \ , r=1,...5$. Under this
assumption Eq.(\ref{Sensitivity1}) assumes the following form
\begin{equation}
\gamma (f) = 
\sqrt{\frac{\sum_{r=1}^5 \lambda^2_r}
{\sum_{i,j=1}^5 \lambda_i \lambda_j \Theta^{ij}}} \
\frac{\sqrt{P_\zeta (f)}}{\rho(f)} \ .
\label{Sensitivity2}
\end{equation}
If we denote with ($\Theta_{min} , \Theta_{max}$) the minimum and
maximum values of the Hellings-Downs curve \cite{HD1983} when
$i \neq j$, and use the identity
$\sum_{r=1}^5 \lambda^2_r = - \sum_{i \neq j = 1}^5 \lambda_i
\lambda_j$
(which follows from the first equation fulfilled by $\vec \lambda$ in
Eq.(\ref{NullSpace2})), from Eq. (\ref{Sensitivity2}) it is then
possible to derive the following inequality
\begin{equation}
\gamma_{max} > \gamma (f)  > \gamma_{min} \ \ , \ \
\gamma_{(min,max)} \equiv \frac{1}{\sqrt{1 - \Theta_{(min,max)}}} \ \frac{\sqrt{P_\zeta
    (f)}}{\rho(f)}  \ .
\label{Sensitivity3}
\end{equation}
If we now multiply and divide the right-hand-side of
Eq.(\ref{Sensitivity3}) by the square-root of the one-sided power
spectral density of the noise of each timing residual, $P_R(f)$, we
derive the following upper-limit for the function $\gamma(f)$
\begin{equation}
\gamma (f)  < \gamma_{max} = \frac{1}{\sqrt{1 - \Theta_{max}}} \
\left(\sqrt{\frac{P_\zeta(f)}{P_R(f)}}\right) \
\left(\frac{\sqrt{P_R(f)}}{\rho(f)}\right) \ .
\label{Sensitivity4}
\end{equation}
Since $\Theta_{min} \simeq -0.15$ and $\Theta_{max} \simeq 0.5$
\cite{HD1983}, the factor that determines the sensitivity gain of the
clock- and ephemeris-free combination $I(t)$ over that of a single
pulsar is the ratio $\sqrt{\frac{P_\zeta(f)}{P_R(f)}}$. This function
of the Fourier frequency can be significantly smaller than one in the
lower part of the band where clock and ephemeris noises dominate the
noise budget. Based on the noise-model discussed in \cite{JAT2011}, in
Fig. \ref{fig2} we plot the estimated GW sensitivity of single-pulsar
experiments together with the sensitivity bounds achievable by the
noise-canceling combination $I(t)$ given in Eq. (\ref{Sensitivity3}).
In this figure we have assumed (i) use of multiple-frequency
measurements to adequately calibrate timing fluctuations from
intergalactic and interplanetary plasma, and (ii) disregarded the
timing fluctuations due to the pulsars.  Stability analysis of known
millisecond pulsars \cite{Verbiest2009,Liu} have shown that there
exist some displaying frequency stabilities superior to those of the
most stable operational clocks in the ($10^{-9} - 10^{-8}$) Hz
frequency band.

As a final remark, arrays with $M \ge 5$ pulsars are characterized by
Kernels of dimensionality equal to $M-4$ (as they have $M-4$
generators) and four-dimensional Images. When $M > 5$ in particular,
the sensitivity of single-source searches can be coherently improved
over that with five pulsars by diagonalizing the correlation matrix
associated with the resulting $M-4$ combinations that are clock- and
ephemeris- free. The anticipated sensitivity enhancement is somewhat
larger than $\sqrt{M-4}$, and we refer the reader to
\cite{Princeetal2002} for details.

\section{Sensitivity to an Isotropic Stochastic GW Background}
\label{SecIV}

The technique presented in this article for canceling correlated
noises (such as clock and ephemeris) affecting data from pulsar timing
arrays becomes particularly effective when searching for a stochastic
background of gravitational radiation with an array of at least $10$
pulsars. This is because arrays with $10$ or more pulsars can generate
pairs of clock- and ephemeris-free combinations whose noises are
expected to be uncorrelated by not sharing data from the same
pulsars. Since the kernel of clock- and ephemeris- free combinations
generated by $M$ pulsars has dimensionality equal to $M-4$, we need to
determine the number of pairs of combinations that can be constructed
with the $M-4$ generators of the kernel and whose noises are
uncorrelated.

To better understand the problem, let us first consider the case of
$10$ pulsars. The associated kernel space is defined by $6$
generators, i.e. any clock- and ephemeris-free combination can be
written as a linear combination of them. If we label the pulsars as
($1, 2, \dots 10$), we can choose, for instance, the generators
constructed by combining the following six set of $5$ pulsars
$(1,2,3,4,5),(6,7,8,9,10),(2,3,4,5,6),(7,8,9,10,1),
(3,4,5,6,7),(8,9,10,1,2)$,
which were obtained by ``circularly-right-shifting'' to the right the
($1, 2, \dots 10$) indices.  With these generators we can only form
the following three pairs of combinations whose noises are
uncorrelated, $[(1,2,3,4,5),(6,7,8,9,10)]$ ;
$[(2,3,4,5,6),(7,8,9,10,1)]$ ; $[(3,4,5,6,7),(8,9,10,1,2)]$, and
implement with them the correlation statistic to search for a GW
stochastic background.

Let us consider one of the above three pairs and denote with
(${}^{(a)}I(t) , {}^{(a)}{\bar I}(t)$) the corresponding two clock-
and ephemeris-free combinations \footnote{Latin indices
  ($a, b, c,\dots h$) will be used to label pairs of clock- and
  ephemeris-free data combinations.}. Here ${}^{(a)}{\bar I}(t)$
contains TOA residuals that are {\underline {not}} entering in the
${}^{(a)}I(t)$ combination and, from Eq. (\ref{ClokEphemerisfree}),
both combinations can be written in the following forms
\begin{eqnarray}
{}^{(a)}I(t) & \equiv &  {}^{(a)}H (t) + {}^{(a)}\zeta (t) = \sum_{j=1}^5 {}^{(a)}\lambda_j {}^{(a)}H^{(j)} + \sum_{j=1}^5
                    {}^{(a)}\lambda_j {}^{(a)}\zeta^{(j)} (t)  \ ,
\nonumber
\\
{}^{(a)}{\bar I}(t) & \equiv &  {}^{(a)}{\bar H} (t) + {}^{(a)}{\bar \zeta} (t) = \sum_{j=1}^5
                       {}^{(a)}{\bar \lambda}_j {}^{(a)}{\bar H}^{(j)} + \sum_{j=1}^5
                       {}^{(a)}{\bar \lambda}_j {}^{(a)}{\bar \zeta}^{(j)} (t)  \ .
\label{Combos}
\end{eqnarray}
Since the noises in these two TOA residual combinations are
uncorrelated, we can implement with them the cross-correlation
statistic \cite{WainsteinZubakov,AllenRomano1999} to search for a
stochastic GW background. To keep as close as possible to the
literature on the implementation of the correlation statistic for GW
searches, we will formulate it in the Fourier domain \footnote{
  Although TOA residuals are not sampled at even rates, use of
  fractional-delay filtering \cite{FDF1,FDF2} allows us to resample
  the data by ``interpolating'' the needed samples at an exquisite level
  of accuracy.}. In what follows we will assume a GW background that is
an isotropic, unpolarized, stationary, and Gaussian random process
with zero mean. Such a background can be characterized by a one-sided
power spectral density, $P_h (|f|)$, defined through the following
expressions \cite{AllenRomano1999,Papoulis2002}
\begin{eqnarray}
\langle {}^{(a)}{\widetilde H}^{(i)} (f) \ {}^{(a)}{\widetilde H}^{*(j)} (f')
  \rangle & \equiv &
\frac{1}{2} \delta (f - f') {}^{(a)} \Theta^{ij} \ P_h (|f|) \ ,
\nonumber
\\
\langle {}^{(a)}{\widetilde {\bar H}}^{(i)} (f) \ {}^{(a)}{\widetilde {\bar H}}^{*(j)} (f')
  \rangle & \equiv &
\frac{1}{2} \delta (f - f') {}^{(a)} {\bar \Theta}^{ij} \ P_h (|f|) \ ,
\nonumber
\\
\langle {}^{(a)} {\widetilde H}^{(i)} (f) \ {}^{(a)} {\widetilde {\bar H}}^{*(j)} (f')
  \rangle & \equiv &
\frac{1}{2} \delta (f - f') {}^{(a)} \Gamma^{ij} \ P_h (|f|) \ ,
\label{HSpectra}
\end{eqnarray}
where angle-brackets, $\langle \rangle$, now denote the operation of
ensemble averaging of the random process and averaging over the
celestial sphere and polarization states, and the three coefficients
(${}^{(a)} \Theta^{ij}, {}^{(a)} {\bar \Theta}^{ij}, {}^{(a)}
\Gamma^{ij}$)
are the Hellings-Downs functions \cite{HD1983} associated with the
correlation of pairs of timing residuals in ${}^{(a)}I(t)$,
${}^{(a)}{\bar I} (t)$, and from both combinations respectively
\footnote{Here ${}^{(a)} \Gamma^{ii}$ is not equal to $1$ as
  it corresponds to the correlation of two different pulsars}. Note
that the spectrum $P_h (|f|)$ appearing in Eq. (\ref{HSpectra})
depends on the spectrum of the GW background, $\Omega_{gw} (f)$, as
well as on the r.m.s. transfer function, $\rho (f)$, of the GW
background to the TOA residual response in the following way
\cite{AllenRomano1999}
\begin{equation}
P_h (|f|) = \rho^2 (f) \ \frac{3 H_0^2}{32 \pi^3} |f|^{-3} \Omega_{gw}(f) \ ,
\label{Ph}
\end{equation}
where $H_0$ is the Hubble constant (today). The typical functional
form for the GW spectrum is a power law, i.e.
$\Omega_{gw} (f) \equiv \Omega_\alpha f^\alpha$.  The spectral index
$\alpha$ characterizes the shape of the GW spectrum and is the unknown
to be determined by the filtering procedure described below.

We will further assume the random processes ${}^{(a)}\zeta^{(i)}$,
${}^{(a)}{\bar \zeta}^{(j)}$ to be stationary, Gaussian distributed
with zero-mean, and characterized by the one-sided power spectral
densities,
($P_{{}^{(a)}\zeta^{(j)}}(|f|) \ , \ P_{{}^{(a)}{\bar
    \zeta}^{(j)}}(|f|)$), defined as follows
\begin{eqnarray}
\langle {}^{(a)} {\widetilde \zeta}^{(i)} (f) \ {}^{(a)} {\widetilde \zeta}^{*(j)} (f')
  \rangle & \equiv & \frac{1}{2} \delta^{ij} \ \delta (f - f') \ P_{{}^{(a)}\zeta^{(i)}} (|f|) 
\nonumber
\\
\langle {}^{(a)} {\widetilde {\bar \zeta}}^{(i)} (f) \ {}^{(a)} {\widetilde {\bar \zeta}}^{*(j)} (f')
  \rangle & \equiv & \frac{1}{2} \delta^{ij} \ \delta (f - f') \
                     P_{{}^{(a)}{\bar \zeta}^{(i)}} (|f|) 
\nonumber
\\
\langle {}^{(a)} {\widetilde \zeta}^{(i)} (f) \ {}^{(a)} {\widetilde {\bar \zeta}}^{*(j)} (f')
  \rangle & \equiv & 0 \ ,
\label{NSpectra}
\end{eqnarray}
where $\delta^{ij}$ is the Kronecker symbol and $\delta (f - f')$ is
the Dirac's delta function. 

The sample cross-correlation statistic is defined as
\begin{equation}
{}^{(a)} {\mathcal S} \equiv \int_{-T/2}^{T/2} dt \ \int_{-\infty}^{\infty} dt'
\ {}^{(a)}I(t) {}^{(a)}{\bar I}(t')
{}^{(a)} Q(t-t') \ ,
\label{S}
\end{equation}
where ${}^{(a)} Q(t-t')$ is the optimal filter that is non-zero over the
time interval ($-T/2, T/2$) (a property we have already taken into account
in Eq.(\ref{S}) by extending the $t'$ integration over the entire real
axis), and reflects the stationarity of both the stochastic background
and the instrumental noises by depending on time differences
\cite{AllenRomano1999}.

From the statistical properties of both the GW stochastic background
and the noises, and by virtue of the central-limit theorem, it follows
that ${}^{(a)} {\mathcal S}$ is also a Gaussian random process
\cite{Papoulis2002} that can be fully characterized by estimating its
mean, ${}^{(a)} \mu_{\mathcal S}$, and variance,
${}^{(a)} \sigma_{\mathcal S}^2$.  Although their derivations are long
(see appendix (\ref{appendixA}) for details), they are straightforward
and result in the following expressions
\begin{equation}
{}^{(a)} \mu_{\mathcal S} \equiv \frac{T}{2} \int^{+\infty}_{-\infty} {}^{(a)}\Lambda(|f|) \ {}^{(a)} {\widetilde Q} (f) \ df 
= \frac{T}{2} \int^{+\infty}_{-\infty} \sum_{r,s=1}^5 {}^{(a)}\lambda_r
  {}^{(a)}{\bar \lambda}_s {}^{(a)} \Gamma^{rs} P_h (|f|) \ {}^{(a)} {\widetilde Q} (f)
\ df \ ,
\label{mus}
\end{equation}
\begin{eqnarray}
{}^{(a)} \sigma_{\mathcal S}^2 & \equiv & \frac{T}{4} \int^{+\infty}_{-\infty} \Delta(|f|) \ |{}^{(a)} {\widetilde Q} (f)|^2 \ df
\nonumber
\\ 
& = & \frac{T}{4} \int^{+\infty}_{-\infty} 
\left[ 
\sum_{r,s,p,q=1}^5 [ {}^{(a)}\lambda_r {}^{(a)}\lambda_s {}^{(a)}{\bar \lambda}_p {}^{(a)}{\bar \lambda}_q  \
      {}^{(a)} \Theta^{rs} \ {}^{(a)} {\bar \Theta}^{pq} 
+ {}^{(a)}\lambda_r {}^{(a)}{\bar \lambda}_s {}^{(a)}\lambda_p {}^{(a)}{\bar \lambda}_q 
\ {}^{(a)} \Gamma^{rs} \ {}^{(a)} \Gamma^{pq} ] \ P^2_h (|f|) \right.
\nonumber
\\
& + & \left. \sum_{r,s,q=1}^5 [ {}^{(a)}\lambda_r {}^{(a)}\lambda_s
      {}^{(a)}{\bar \lambda}^2_q \ {}^{(a)} \Theta^{rs} \ P_{{}^{(a)}{\bar \zeta}^{(q)}}(|f|) +
{}^{(a)}{\bar \lambda}_r {}^{(a)}{\bar \lambda}_s {}^{(a)}\lambda^2_q
     \ {}^{(a)} {\bar \Theta}^{rs} \ P_{{}^{(a)}\zeta^{(q)}}(|f|) ] \ P_h (|f|) \right.
\nonumber
\\
& + & \left. \sum_{r,s=1}^5 {}^{(a)}\lambda_r^2 \ {}^{(a)}{\bar \lambda}^2_s
         \ P_{{}^{(a)}\zeta^{(r)}}(|f|) \ P_{{}^{(a)}{\bar \zeta}^{(s)}}(|f|) \right] \ |{}^{(a)} {\widetilde Q} (f)|^2 \ df \ .
\label{sigmas}
\end{eqnarray}
By simple inspection of Eq. (\ref{sigmas}), and following
\cite{AllenRomano1999}, it is convenient to define the following
operation of inner product between two arbitrary complex functions,
say $A$, and $B$
\begin{equation}
(A,B) \equiv \int^{+\infty}_{-\infty} A(f) B^* (f) \Delta(|f|) \ df \ .
\label{inner}
\end{equation}
With this newly defined inner product, ${}^{(a)} \mu_{\mathcal S}$ and
${}^{(a)} \sigma_{\mathcal S}$ can be
rewritten in the following forms
\begin{equation}
{}^{(a)} \mu_{\mathcal S} = \frac{T}{2} ({}^{(a)} {\widetilde Q}, \frac{{}^{(a)} \Lambda}{{}^{(a)} \Delta}) \ \ \
\ ; \ \ \ \ {}^{(a)} \sigma^2_{\mathcal S} = \frac{T}{4} ({}^{(a)}
{\widetilde Q} (f) , {}^{(a)} {\widetilde Q} (f) ) \ ,
\end{equation}
while the squared signal-to-noise ratio of ${}^{(a)} {\mathcal S}$,
${}^{(a)} SNR^2$, is equal to
\begin{equation}
{}^{(a)} SNR^2 \equiv \frac{{}^{(a)} \mu_{\mathcal S}^2}{{}^{(a)}
  \sigma_{\mathcal S}^2} = T \ 
\frac{ ({}^{(a)} {\widetilde Q}, \frac{{}^{(a)} \Lambda}{{}^{(a)} \Delta})^2}{({}^{(a)} {\widetilde Q} , {}^{(a)} {\widetilde Q} )} \ .
\label{SNRP}
\end{equation}
From a simple geometrical interpretation of the inner product defined
through Eq. (\ref{inner}), from Eq. (\ref{SNRP}) we conclude that the
maximum of the SNR of ${}^{(a)} {\mathcal S}$ is achieved by choosing
the filter function ${}^{(a)} {\widetilde Q}$ to be equal to
\begin{equation}
{}^{(a)} {\widetilde Q} (f) = {}^{(a)} {\xi} \ \frac{{}^{(a)} \Lambda}{{}^{(a)} \Delta} \ ,
\label{Q}
\end{equation}
where ${}^{(a)} {\xi}$ is an arbitrary real number and, from the
definition of the functions ${}^{(a)} \Lambda$ and ${}^{(a)} \Delta$ (Eqs. \ref{mus},
\ref{sigmas}), it also follows that ${}^{(a)} {\widetilde Q} (f)$ is
real. We can take advantage of the arbitrariness of ${}^{(a)} {\xi}$
by making all mean values of the cross-correlations equal to the
following constant $\mu_{\mathcal S} \equiv {}^{(a)} \mu_{\mathcal S} = \Omega_\alpha \ T \ , \ a = 1, 2, 3$
\cite{AllenRomano1999}. As it will become clearer later on in this
section, this choice simplifies the derivation of the coherent SNR
achievable by properly combining the cross-correlations of pairs of
clock- and ephemeris-free combinations.

The expression for the filter ${}^{(a)} {\widetilde Q}$ given by Eq. (\ref{Q})
implies the following maximum SNR achievable by cross-correlating the
pair (${}^{(a)}I(t) , {}^{(a)}{\bar I}(t)$)
\begin{equation}
{}^{(a)} SNR_{{}^{(a)}I} \equiv \sqrt{T} \ \sqrt{ \int^{+\infty}_{-\infty}
\frac{{}^{(a)} \Lambda^2(|f|)}{{}^{(a)} \Delta(|f|)} \ df } \ .
\label{SNRI}
\end{equation}
In the limit in which the noise spectra of the TOA residuals are
larger than that of the GW background, the integrand of
Eq.(\ref{SNRI}) becomes equal to
\begin{equation}
\frac{{}^{(a)} \Lambda^2(|f|)}{{}^{(a)} \Delta(|f|)} \simeq 
\frac{ \left( \sum_{r,s=1}^5 {}^{(a)}\lambda_r {}^{(a)}{\bar \lambda}_s \Gamma^{rs} \right)^2 \
P^2_h (|f|)}
{ \left( \sum_{r=1}^5 {}^{(a)}\lambda^2_r P_{{}^{(a)}\zeta^{(r)}}(|f|) \right) \
 \left( \sum_{s=1}^5 {}^{(a)}{\bar \lambda}^2_s P_{{}^{(a)}{\bar \zeta}^{(s)}}(|f|)
 \right) } \ .
\label{IntegrandI}
\end{equation}
Note that the above expression of the optimal SNR depends on the
pulsars' relative sky locations through the Hellings and Downs
correlation function and the vectors,
${{}^{(a)}{\vec \lambda}} , {{}^{(a)}{\bar {\vec \lambda}}}$
identifying the pulsars' clock- and ephemeris-free combinations. To
quantify the angular dependence of the optimal SNR, we can assume the
timing residual noises
(${}^{(a)}\zeta^{(r)}, {}^{(a)} {\bar {\zeta}}^{(r)} \ , \ r = 1,
\dots 5$)
to be characterized by the same spectrum, $P_\zeta (|f|)$. Under this
assumption the optimal SNR becomes equal to
\begin{equation}
{}^{(a)} SNR_{{}^{(a)}I} \equiv \sqrt{T} \ 
\frac{ | \sum_{r,s=1}^5 {}^{(a)}\lambda_r {}^{(a)}{\bar \lambda}_s
  \Gamma^{rs} | }{\sqrt{\sum_{r,s=1}^5 {}^{(a)}\lambda^2_r
    {}^{(a)}{\bar \lambda}^2_s}} \sqrt{ \int^{+\infty}_{-\infty}
  \frac{P^2_h (|f|)}{P_{\zeta}^2 (|f|)} \ df } \ , 
\label{SNRI_2}
\end{equation}
where the dependence of the SNR on the pulsars' relative sky locations
can now be factored out of the integral. To quantify the magnitude of
the angular function appearing in Eq. (\ref{SNRI_2}) we have randomly
generated $10^7$ sets of $10$ pulsars' sky locations and compute it
for each set. We found the above angular function to assume values
within the interval ($0, 1$) and to have an r.m.s. value equal to
about $1/5$.

To compare the effectiveness of our cross-correlation statistic
against that based on a pair of TOA residuals,
($R^{(i)} (t) , R^{(j)} (t)$), we provide below the expression for the
optimal SNR, $SNR^{ij}_{R}$, associated with the cross-correlation
statistic of two timing residuals \footnote{The expression of the
  optimal SNR of the cross-correlation statistic of a pair of timing
  residuals follows from their definitions (Eq. (\ref{Residuals})),
  and by performing a calculation similar to that presented in
  Appendix \ref{appendixA}.}
\begin{equation}
SNR^{ij}_{R} = \sqrt{T}   \sqrt{ \int^{+\infty}_{-\infty} df \ 
\frac{ 
[ \Gamma^{ij} P_h(|f|) + P_C(|f|) + P_{e^{ij}}(|f|) ]^2
}{\Pi (|f|)} } \ ,
\label{SNRR_Limit}
\end{equation}
where:
\begin{eqnarray}
\Pi (|f|) & \equiv& 2P^2_C(|f|) + P_{e^{ii}} (|f|) P_{e^{jj}} (|f|) +
  P^2_{e^{ij}}(|f|) + [P_{e^{ii}} (|f|)  + P_{e^{jj}} (|f|) ] P_\zeta(|f|)  
\nonumber
\\
& + & [P_{e^{ii}} (|f|)  + P_{e^{jj}} (|f|)  + 2
  P_{e^{ij}} (|f|) + 2 P_\zeta (|f|) ] P_C(|f|) + P^2_{\zeta}(|f|) \ .
\label{PI}
\end{eqnarray}
Eqs. (\ref{SNRR_Limit}, \ref{PI}) reflect the assumption on the
spectra of the $\zeta$-noises to be equal to each other and larger
than the GW background, and where we have also denoted with
($P_C(|f|) , P_{e^{ij}}(|f|)$) the spectra of the clock and ephemeris
noises respectively. Eqs. (\ref{SNRR_Limit}, \ref{PI}) allow us to
quantify two aspects of the degradation in the likelihood of detection
due to correlated-noises in pulsar timing data
\cite{Coles2011,Tiburzi2015,NANOGRAV1}. First, the cross-correlation
statistic of pairs of TOA residuals from different pulsars is affected
by clock and ephemeris noises through their contribution to the mean
value of the cross-correlation (i.e. the numerator of the integrand in
Eq. (\ref{SNRR_Limit})). Since these noises are characterized by
relatively large spectral components in the same part of the band
associated with the presence of a stochastic GW background, they
result in an increased {\it false-alarm probability} . Second, clock
and ephemeris noises contribute to the overall noise variance of the
TOA residuals and therefore reduce (by a factor larger than $10$
\cite{JAT2011}) the optimal cross-correlation SNR
(Eq. (\ref{SNRR_Limit})) over that associated with pairs of clock and
ephemeris-free combinations (Eq. (\ref{SNRI_2})).

Following \cite{AllenRomano1999}, we now provide the expression for
the SNR achievable by optimally combining the three pairs of clock-
and ephemeris-free combinations that can be synthesized with the
timing data from an array of $10$ pulsars. This requires the
calculation of the inverse of the variance-covariance matrix of the
cross-correlation statistic,
${}^{(ab)} C \equiv \langle {}^{(a)} {\mathcal S} {}^{(b)} {\mathcal
  S} \rangle - \langle {}^{(a)} {\mathcal S} \rangle \ \langle
{}^{(b)} {\mathcal S} \rangle$.
The discussion on how to derive ${}^{(ab)} C$ is presented in the
appendix, and its expression can be written in the following form
\begin{eqnarray}
{}^{(ab)} C & = & \frac{T}{4} \int^{+\infty}_{-\infty} 
\left[ \sum_{r,s,p,q=1}^5 [ {}^{(a)}\lambda_r {}^{(b)}\lambda_s
      {}^{(a)}{\bar \lambda}_p {}^{(b)}{\bar \lambda}_q  
      {}^{(ab)} \Theta^{rs} {}^{(ab)} {\bar \Theta}^{pq} + 
{}^{(a)}\lambda_r {}^{(b)}{\bar \lambda}_s {}^{(a)}{\bar \lambda}_p {}^{(b)}\lambda_q
{}^{(ab)} \Gamma^{rs} {}^{(ab)} {\bar \Gamma}^{pq} ] \ P^2_h (|f|)
\right.
\nonumber
\\
& + & \left. \sum_{r,s,p,q=1}^5  [ {}^{(a)}\lambda_r {}^{(b)}\lambda_s
      {}^{(a)}{\bar \lambda}_p {}^{(b)}{\bar \lambda}_q {}^{(ab)} \Theta^{rs} \eta^{(ab)}_{pq}
\ {}^{(ab)} P_{{\bar \zeta}_p}(|f|) +
{}^{(a)}\lambda_r {}^{(b)}\lambda_s
      {}^{(a)}{\bar \lambda}_p {}^{(b)}{\bar \lambda}_q {}^{(ab)}{\bar \Theta}^{pq} 
\eta^{(ab)}_{rs} \ {}^{(ab)} P_{{\zeta}_r}(|f|) 
\right.
\nonumber
\\
& + & \left. {}^{(a)}\lambda_r  {}^{(b)}{\bar \lambda}_s {}^{(a)}{\bar \lambda}_p
      {}^{(b)}\lambda_q 
{}^{(ab)} \Gamma^{rs} \eta^{(ab)}_{pq} \ {}^{(ab)} P_{{\zeta}_p} (|f|) +
{}^{(a)}\lambda_r  {}^{(b)}{\bar \lambda}_s {}^{(a)}{\bar \lambda}_p
      {}^{(b)}\lambda_q 
{}^{(ab)} \Gamma^{pq} \eta^{(ab)}_{rs} \ {}^{(ab)} P_{{\bar \zeta}_r}(|f|) ] \ P_h (|f|) 
\right.
\nonumber
\\
& + & \left. \sum_{r,s,p,q=1}^5 {}^{(a)}\lambda_r  {}^{(b)}{\bar \lambda}_s {}^{(a)}{\bar \lambda}_p
      {}^{(b)}\lambda_q \eta^{(ab)}_{rs} \eta^{(ab)}_{pq} \ {}^{(ab)}
      P_{{\bar \zeta}_r}(|f|) \ {}^{(ab)} P_{{\zeta}_p}(|f|) \right] \  {}^{(a)} {\widetilde Q} (f) {}^{(b)} {\widetilde Q}^* (f)  \ df \ .
\label{Cab}
\end{eqnarray}
In the above equation the multi-indices symbol $\eta^{(ab)}_{pq}$ is
either equal to $0$ or $1$ depending on the particular outcome of the
correlation of two $\zeta$ noises entering in the combinations' pair
$(ab)$, and it reduces to $\delta_{pq}$ when $a=b$.  The expression
for the optimal SNR achievable by combining the cross-correlations of
our three pairs of clock- and ephemeris-free combinations is equal to
\cite{AllenRomano1999}
\begin{equation}
SNR_{opt}^2 = ({}^{(1)} \mu_{\mathcal S}, {}^{(2)} \mu_{\mathcal S}, {}^{(3)} \mu_{\mathcal S})
\ C^{-1} \ 
({}^{(1)} \mu_{\mathcal S}, {}^{(2)} \mu_{\mathcal S}, {}^{(3)}
\mu_{\mathcal S})^{\mathcal T}  = (\Omega_\alpha T)^2 \ \sum_{a,b = 1}^{3} {}^{(ab)} (C^{-1}) \ ,
\label{SNRTotal}
\end{equation}
where the mean values ${}^{(i)} \mu_{\mathcal S} \ , \ i=1, 2, 3$ have
been normalized to the same constant value $\Omega_\alpha T$.

The analysis for a set of $10$ pulsars presented above can be
generalized to an arbitrary array of size $M$. This is done by first
selecting the $M-4$ generators of the kernel, and then identifying the
set of all pairs of generators that do not have pulsars in common,
i.e. pairs of clock- and ephemeris-free combinations whose noises are
uncorrelated. As an example of how to identify such set of generators'
pairs, let us consider an array with $45$ pulsars, equal in number to
the array recently analyzed by the NANOGRAV consortium
\cite{NANOGRAV2}.  Let us first consider the numerical vector
($1, 2, \dots 45$), which can be used to label the $45$ pulsars. If we
use the ``right-circular-shifting'' procedure described earlier for
selecting a set of generators of the $10$-pulsar kernel, we obtain the
following $41$ generators
\begin{eqnarray}
& ( & 1,\dots 5) , (6, \dots 10) \dots (41, \dots 45) 
\nonumber
\\
& ( & 45, \dots 4) , (5, \dots 9) \dots (40, \dots 44) 
\nonumber
\\
& ( &44, \dots 3) , (4, \dots 8) \dots (39, \dots 43) 
\nonumber
\\
& ( & 43, \dots 2) , (3, \dots 7) \dots (38, \dots 42) 
\nonumber
\\
& ( & 42, \dots 1) , (2, \dots 6) , (7, \dots 11) , (12, \dots 16) , (17, \dots 21) 
\label{fortyone}
\end{eqnarray}
This set of $41$ clock- and ephemeris-free combinations of timing
residuals results in a total of $672$ pairs that do not share data
from the same pulsars, and can therefore be used to implement the
cross-correlation statistic\footnote{The total number of $672$ pairs
  of clock- and ephemeris-free combinations generated by an array of
  $45$ pulsars was calculated numerically by using the program {\it
    Mathematica} \cite{Mathematica}}. Since this number is comparable
to the number of pairs ($990$) of timing residuals given by an array
$45$ pulsars (and upon which the usual cross-correlation statistic is
built), it follows that the optimal SNRs associated with both
cross-correlation statistics will scale roughly by the same amount
over their respective single-pair SNRs \cite{AllenRomano1999}.

\section{Conclusions}
\label{SecV}

The data processing technique presented in this article allows us to
cancel clock and ephemeris noises affecting nHz GW pulsar timing
experiments. This is done by properly constructing linear combinations
of TOA residuals generated by arrays of millisecond pulsars.

We have found that searches for single-source GW signals will benefit
from this technique when implemented with arrays of at least $5$
millisecond pulsars. The estimated sensitivity enhancement over that
from individual pulsar experiments is of at least one order of
magnitude in the lower-part of the accessible frequency band. This is
the frequency region where clock and ephemeris noises are leading
noise sources equally effecting the array's timing measurements.

Searches for an isotropic stochastic GW background can also be
performed with clock- and ephemeris-free combinations from an array of
$10$ or more pulsars. This is done by implementing the
cross-correlation statistic with pairs of clock- and ephemeris-free
combinations that do not share timing residuals from the same pulsars
to prevent noise correlations. With an array of $10$ pulsars we have
found the associated cross-correlation statistic to be characterized
by an optimal SNR that is more than an order of magnitude larger than
the optimal SNR achievable by cross-correlating pairs of timing
residuals from an equal-size array.

As a final note, we have shown that clock and ephemeris noises can be
reconstructed with the timing data from an array of $4$ or more
pulsars. In a future article we will estimate the accuracies by which
these observable can be reconstructed as functions of the number of
pulsars, their relative sky locations, and the magnitudes of the
remaining noises affecting the timing measurements.

\section*{Acknowledgments}

I would like to thank Dr. Stephen Taylor for the many useful
conversations during the development of this work, Dr. Lee Lindblom
for several comments during the early development phase of the idea
presented in this article, and Dr. Frank B. Estabrook and John
W. Armstrong for their constant encouragement.

\appendix
\section{Derivation of the mean and variance-covariance of ${\mathcal S}$}
\label{appendixA}

In what follows we derive the expressions of the mean and the
variance-covariance matrix of the cross-correlation function
${}^{(a)} S$ (Eqs. (\ref{mus}, \ref{sigmas}, \ref{Cab}) given in
Section (\ref{SecIV})) for an array of $10$ pulsars. We will assume
the noises $\zeta_r$ to be Gaussian-distributed with zero-mean, and to
have one-sided power-spectral densities $P_{\zeta_r}(|f|)$. We will
also take the GW stochastic background to be a Gaussian random process
of zero-mean, unpolarized, and stationary.  As a consequence of these
assumptions such a background is uniquely characterized by a one-sided
power spectral density, $P_h (|f|)$ as given by Eqs.  (\ref{HSpectra},
\ref{Ph}).

From the expressions of the clock- and ephemeris-free data
combinations, ${}^{(a)} I(t)$, their complements,
${}^{(a)} {\bar I} (t)$, (i.e. those combinations that do not include
timing data from the pulsars entering in ${}^{(a)} I(t)$), and the
definition of their cross-correlation statistic,
${}^{(a)} {\mathcal S}$, we have
\begin{equation} {}^{(a)} {\mathcal S} \equiv \int_{-T/2}^{T/2} dt \
  \int_{-\infty}^{\infty} dt' \ {}^{(a)}I(t) {}^{(a)}{\bar I}(t')
  {}^{(a)} Q(t-t') \ ,
\label{A_S}
\end{equation}
where ${}^{(a)} Q(t-t')$ is the optimal filter. Since this is non-zero
over the interval ($-T/2, T/2$), we have extended the $t'$ integration
over the entire real axis. Equivalently, Eq. (\ref{A_S}) can be
rewritten in the Fourier domain as follows
\begin{equation}
{}^{(a)} {\mathcal S} = \int_{-\infty}^{\infty} df \ \int_{-\infty}^{\infty} df'
\ \delta_T (f - f') \ {}^{(a)}{\widetilde I}^* (f) {}^{(a)} {\widetilde {\bar I}}(f')
{}^{(a)} {\widetilde Q} (f') \ ,
\label{A_FS}
\end{equation}
where $\delta_T (f) \equiv T sinc(\pi f T)$ is the finite-time
approximation of the Dirac's delta function.

Since the noises in the combination ${}^{(a)} I(t)$ do not enter in
${}^{(a)} {\bar I} (t)$, and because a GW stochastic background is
uncorrelated with the measurement noises, we infer that the ensemble
average of the cross-correlation statistic, Eq. (\ref{A_FS}), contains
contribution only from the GW stochastic background in the following
form
\begin{eqnarray}
{}^{(a)} \mu_{\mathcal S} \equiv \langle {}^{(a)} {\mathcal S} \rangle
& = & \int_{-\infty}^{\infty} df \ \int_{-\infty}^{\infty} df'
\ \delta_T (f - f') \ \langle {}^{(a)}{\widetilde I}^* (f) {}^{(a)} {\widetilde {\bar
    I}}(f') \rangle {}^{(a)} {\widetilde Q} (f') 
\nonumber
\\
& = &
\int_{-\infty}^{\infty} df \ \int_{-\infty}^{\infty} df'
\ \delta_T (f - f') \ \langle {}^{(a)} {\widetilde H}^* (f) {}^{(a)} {\widetilde {\bar
    H}}(f') \rangle {}^{(a)} {\widetilde Q} (f') \ ,
\label{Amu1}
\end{eqnarray}
where ${}^{(a)}{\widetilde H}^* (f)$, $ {}^{(a)} {\widetilde {\bar H}}(f')$
are the contributions of the GW stochastic background to the
clock- and ephemeris-free combinations ${}^{(a)}{\widetilde I}^* (f)$,
${}^{(a)} {\widetilde {\bar I}}(f')$ respectively. Since
${}^{(a)}{\widetilde H}^* (f) = \sum_{r=1}^5 {}^{(a)}\lambda_r
{}^{(a)}{\widetilde H}^{*r} (f)$, and 
${}^{(a)} {\widetilde {\bar H}} (f') = \sum_{r=1}^5 {}^{(a)} {\bar \lambda}_r
{}^{(a)}{\widetilde {\bar H^r}} (f)$, we have
\begin{eqnarray}
\langle {}^{(a)} {\widetilde H}^* (f) {}^{(a)} {\widetilde {\bar
  H}}(f') \rangle & = &
\sum_{r,s=1}^5 {}^{(a)}\lambda_r {}^{(a)} {\bar \lambda}_s \langle {}^{(a)} {\widetilde H}^{*r} (f) {}^{(a)}
{\widetilde {\bar H^s}} (f') \rangle 
\nonumber
\\
& = &
\frac{1}{2} \delta(f - f')
\sum_{r,s=1}^5 {}^{(a)}\lambda_r {}^{(a)} {\bar \lambda}_s {}^{(a)}
\Gamma^{rs} \ {}^{(a)}P_h (|f|) \ ,
\label{Amu2}
\end{eqnarray}
where ${}^{(a)} \Gamma^{rs}$ is the Hellings and Downs correlation function
\cite{HD1983}. After substituting Eq. (\ref{Amu2}) into
Eq. (\ref{Amu1}) and exercising the Dirac's delta function, we finally
obtain Eq. (\ref{mus}).

From the definition of the variance-covariance matrix of the cross-correlation
statistic,
\begin{equation}
{}^{(ab)} C \equiv \langle {}^{(a)} {\mathcal
  S} {}^{(b)} {\mathcal S} \rangle - \langle {}^{(a)} {\mathcal S}
\rangle \langle {}^{(b)} {\mathcal S} \rangle \ ,
\end{equation}
we have
\begin{eqnarray}
{}^{(ab)} C  & = &
 \int_{-\infty}^{\infty} df \ \int_{-\infty}^{\infty} df'  \int_{-\infty}^{\infty} dk \ \int_{-\infty}^{\infty} dk'
\ \delta_T (f - f') \ \delta_T (k - k') 
\nonumber
\\
& \times &
\ \left[ \langle {}^{(a)}{\widetilde I}^* (f) {}^{(a)} {\widetilde {\bar I}}(f')
\ {}^{(b)}{\widetilde I} (k) {}^{(b)} {\widetilde {\bar I}}^*(k') \rangle - 
\langle {}^{(a)}{\widetilde I}^* (f) {}^{(a)} {\widetilde {\bar I}}(f') \rangle
\langle {}^{(b)}{\widetilde I} (k) {}^{(b)} {\widetilde {\bar
           I}}^*(k') \rangle \right]
\nonumber
\\
& \times & \ {}^{(a)} {\widetilde Q} (f') {}^{(b)} {\widetilde Q}^* (k') \ .
\label{Asigmas1}
\end{eqnarray}
Since the random processes associated with the noises and the GW
background are Gaussian, we infer that the linear combinations
${}^{(a)}{\widetilde I} (f)$, ${}^{(a)} {\widetilde {\bar I}}(f')$,
${}^{(b)}{\widetilde I} (f)$, ${}^{(b)} {\widetilde {\bar I}}(f')$ are
also Gaussian. This implies that the term in the integrand containing
four $I$-combinations, inside the ensemble average operator, can be
written in the following form \cite{Papoulis2002}
\begin{eqnarray}
\langle 
{}^{(a)}{\widetilde I}^* (f) 
{}^{(a)} {\widetilde {\bar I}}(f')
{}^{(b)}{\widetilde I} (k) 
{}^{(b)} {\widetilde {\bar I}}^*(k')
\rangle & = &
\langle 
{}^{(a)}{\widetilde I}^* (f) 
{}^{(a)} {\widetilde {\bar I}}(f') 
\rangle 
\langle 
{}^{(b)}{\widetilde I} (k) 
{}^{(b)} {\widetilde {\bar I}}^*(k') 
\rangle
\nonumber
\\
& + &
\langle 
{}^{(a)}{\widetilde I}^* (f) 
{}^{(b)}{\widetilde I} (k) 
\rangle 
\langle 
{}^{(a)} {\widetilde {\bar I}}(f') 
{}^{(b)} {\widetilde {\bar I}}^*(k') 
\rangle
\nonumber
\\
& + &
\langle 
{}^{(a)}{\widetilde I}^* (f) 
{}^{(b)} {\widetilde {\bar I}}^*(k') 
\rangle 
\langle 
{}^{(a)} {\widetilde {\bar I}}(f') 
{}^{(b)}{\widetilde I} (k) 
\rangle
\label{GaussIdentity}
\end{eqnarray}
After substituting Eq. (\ref{GaussIdentity}) into Eq. (\ref{Asigmas1})
we get
\begin{eqnarray}
{}^{(ab)} C & = &
 \int_{-\infty}^{\infty} df \ \int_{-\infty}^{\infty} df'  \int_{-\infty}^{\infty} dk \ \int_{-\infty}^{\infty} dk'
\ \delta_T (f - f') \ \delta_T (k - k') 
\nonumber
\\
& \times &
\ \left[ 
\langle 
{}^{(a)}{\widetilde I}^* (f) 
{}^{(b)}{\widetilde I} (k) 
\rangle 
\langle 
{}^{(a)} {\widetilde {\bar I}}(f') 
{}^{(b)} {\widetilde {\bar I}}^*(k') 
\rangle
+
\langle 
{}^{(a)}{\widetilde I}^* (f) 
{}^{(b)} {\widetilde {\bar I}}^*(k') 
\rangle 
\langle 
{}^{(a)} {\widetilde {\bar I}}(f') 
{}^{(b)}{\widetilde I} (k) 
\rangle
\right]
\nonumber
\\
& \times & \ {}^{(a)} {\widetilde Q} (f') {}^{(b)} {\widetilde Q}^* (k') \ .
\label{Asigmas2}
\end{eqnarray}
By replacing the $I$s in terms of their GW signal and noise terms, and
applying the ensemble average operation (see Eqs. (\ref{HSpectra},
\ref{NSpectra}) on the resulting expressions entering in
Eq. (\ref{Asigmas2}), Eq. (\ref{Cab}) can be obtained after some
straightforward algebra.
\bibliography{MyRefs}
\end{document}